\documentclass[useAMS,usenatbib]{mn2e}
\usepackage{natbib}
\usepackage{graphicx}
\usepackage{subfigure}
\usepackage{multirow}
\usepackage{amsmath}
\usepackage{amssymb}
\usepackage{epstopdf}
\usepackage{rotating}
\usepackage{soul}
\usepackage{placeins}
\usepackage{color}

\DeclareUnicodeCharacter{2212}{-}

\title{On the detectability of massive black hole merger events by LISA}

\author[Banks et al.]{Samuel Banks$^{1}$\thanks{Undergraduate student},  Katharine Lee$^{1}\footnotemark[1]$, Nazanin Azimi$^{1}\footnotemark[1]$, 
Kendall Scarborough$^{1}\footnotemark[1]$, 
\newauthor Nikolai Stefanov$^{1}\footnotemark[1]$, Indra Periwal$^{1}\footnotemark[1]$, Colin DeGraf$^{2}$\thanks{Contact e-mail:cdegraf@cmu.edu }, Tiziana Di Matteo$^{2,3,4}$ \\
$^{1}$ Department of Physics, 
Carnegie Mellon University, Pittsburgh, PA 15213 \\
$^{2}$ McWilliams Center for Cosmology, Department of Physics, Carnegie Mellon University, Pittsburgh, PA 15213 \\
$^{3}$ NSF AI Planning Institute for Physics of the Future, 
Carnegie Mellon  University, Pittsburgh, PA 15213, USA \\
$^{4}$ OzGrav-Melbourne, Australian Research Council Centre of Excellence for Gravitational Wave Discovery\\
}

\date{}

\begin{document}

\maketitle
\begin{abstract}
The launch of space based gravitational wave (GW) detectors (e.g. Laser Interferometry Space Antenna; LISA) and current and upcoming Pulsar Timing Arrays (PTAs) will extend the GW window to low frequencies, 
opening new investigations into dynamical processes involving massive black hole binaries (MBHBs) and their mergers across cosmic time.  MBHBs are expected to be among the primary sources for the upcoming low frequency  ($10^{−4} − 10^{−1}$ Hz) window probed by LISA.
   It is important to investigate the expected MBH merger rates and associated signals, to determine how potential LISA events are affected by physics included in current models. To study this, 
   we post-process the large population of MBHBs in the Illustris simulation 
to account for dynamical friction time delays associated with BH infall/inspiral.  
   We show that merger delays associated with binary evolution have the potential to decrease the expected merger rates, with $M_{\rm{BH}} > 10^6 M_\odot$ MBHBs (the lowest mass in Illustris) decreasing from $\sim 3$ yr$^{-1}$ to $\sim 0.1 $yr$^{-1}$, and shifting the merger peak from z $\sim 2$ to $\sim 1.25$.
    During this time, we estimate that accretion grows the total merging mass by as much as 7x from the original mass. 
   Importantly, however, dynamical friction associated delays (which shift the mergers toward lower-redshift and higher-masses) lead to a stronger signal/strain for the emitted GWs in the LISA band, increasing mean frequency from $10^{-3.1}$ to $10^{-3.4}-10^{-4.0}$ Hz, and mean strain from $10^{-17.2}$ to $10^{-16.3}-10^{-15.3}$.  Finally, we show that after including a merger delay and associated $M_{\rm{BH}}$ growth, mergers still tend to lie on the typical $M_{\rm{BH}}-M_*$ relation, but with an increased likelihood of an undermassive black hole.

\end{abstract}

\section{INTRODUCTION}

It has been well established that the supermassive black holes found at the center of galaxies \citep[e.g.][]{KormendyRichstone1995} are correlated with the properties of their host galaxies \citep[e.g.][]{Magorrian1998, Gebhardt2000, Graham2001, Ferrarese2002, Tremaine2002, HaringRix2004, Gultekin2009, McConnellMa2013, KormendyHo2013, Reines2015, Greene2016, Schutte2019}.  As dark matter halos and galaxies merge \citep[e.g.][]{Fakhouri2010, Rodriguez-Gomez2015}, we expect the black holes hosted by the merging galaxies to migrate toward the galactic center, where they are able to form binaries and eventually merge themselves \citep[e.g.][]{Mayer2007}.  SMBH mergers are also the strongest sources of gravitational wave emission in the Universe, making them an important source for upcoming surveys. 

In recent years, gravitational wave (GW) signals have been detected from black hole mergers \citep[e.g.][]{Abbott2016}, but to this point the gravitational wave detections have been limited to stellar mass black holes.  Mergers between supermassive black holes produce much longer wavelength GWs, beyond the sensitivity of current ground based interferometers.  However, the upcoming LISA space mission will be capable of detecting longer-wavelength GWs, with peak sensitivity at $\sim 10^4-10^7 M_\odot$ \citep{AmaroSeoane2017}, while Pulsar Timing Arrays (PTAs) should be sensitive to even longer wavelengths, reaching black holes with masses above $10^8 M_\odot$ \citep{Verbiest2016, Desvignes2016, Reardon2016, Arzoumanian2018}.

GW detections are expected to provide extensive information on a wide range of black hole properties, including the rate of supermassive black hole mergers across cosmic time \citep[e.g.][]{Klein2016, Salcido2016, Kelley2017, Ricarte2018, Katz2020, Volonteri2020}, their connection to black hole - galaxy scaling relations \citep[e.g.][]{VolonteriNatarajan2009, Simon2016, Shankar2016}, and how supermassive seeds form \citep[e.g.][]{Sesana2007, Ricarte2018, DeGrafSijacki2020}, while multimessenger studies combining GW and electromagnetic data should provide additional information about the galaxies in which mergers occur \citep[e.g.][]{Volonteri2020, DeGraf2021}.  Before the first such GWs from supermassive black hole mergers are detected, however, it is important to provide predictions for these expected events to maximize what we can learn as these new GW sources are discovered. 

Cosmological hydrodynamical simulations provide an effective tool with which to study these mergers, as they incorporate black holes and galaxy formation, and span sufficient volume to include large statistical samples.  Current simulations \citep[e.g.][]{Vogelsberger2014a, Dubois2014, Schaye2015, Feng2016, Pillepich2018b, Henden2018, Dave2019, Chen2021} contain a wide range of black hole masses (typically between $10^4-10^{10} M_\odot$) and can be used to provide detailed investigations into a wide range of black hole properties, including black hole mergers. 
One of the limitations of current cosmological simulations is the treatment for black hole motion.  There has been significant work toward implementing more physical models for black hole motion within galaxies \citep[e.g.][]{Tremmel2015, Tremmel2017, Pfister2019, Sharma2020, Ricarte2021, Chen2021} and the formation and coalescence of black hole binaries \citep[e.g.][]{Pfister2017, Rantala2017}.  Nonetheless, cosmological simulations commonly use a simpler recentering scheme and instantaneously merge black hole pairs, which has the potential to significantly impact the rate at which GWs are detected \citep[e.g.][]{Katz2020, Volonteri2020, DeGraf2021}.

Recently, some simulations have been analyzed to consider a more accurate estimate for the merging timescales, and the impact on overall merger populations. \citet{Volonteri2020} estimated merging timescales in the HorizonAGN and NewHorizon simulations, using a combination of dynamical friction and binary hardening to delay mergers, and found that these timescales are long enough that black hole mergers occur well after their host galaxy mergers, such that host galaxies no longer have disturbed morphologies, and the black hole merger rates peak at lower redshift than in the original simulation.  Similarly, \citet{Katz2020} calculate binary lifetimes for black hole mergers in the Illustris simulation, and perform Monte Carlo simulations to generate mock merger catalogs to estimate coalescence timescales and merger rates in detail.  

In this paper, we perform a post-processing analysis on the Illustris simulation \citep{Nelson2015} to test the impact of longer merging timescales.  We use a rough estimate for the merger timescale based on dynamical friction within the host galaxy, which we use to delay each individual merger.  We further estimate the expected growth (via gas accretion) of each black hole during this time under assumed Bondi- or Eddington- scaling, to determine the new black hole masses involved in each merger.   We then compare this new set of merger data to the mergers in the original simulation, and quantify the impact on the supermassive black hole merger rates, merging masses, gravitational wave signals (and thus observability by LISA), and the connection to their host galaxies.  

In Section \ref{sec:method} we describe the Illustris simulation and the post processing calculations for the merger delay and associated black hole mass growth.  In Section \ref{sec:populations} we characterize the populations of black hole mergers, looking at both the typical merging masses and the redshifts at which they take place.  In Section \ref{sec:GWdetect} we calculate the frequency and strain of the gravitational waves emitted by each merger and the total rate at which GW signals are expected to reach the Earth; and in Section \ref{sec:scaling} we look at the connection between black hole mergers and their host galaxies, and compare to the general (non-merging) black hole population.  Finally, we summarize our conclusions in Section \ref{sec:conclusions}.

\section{Method}
\label{sec:method}
\subsection{Simulation}

In this work, we use the Illustris\footnote{https://www.illustris-project.org} simulation \citep{Nelson2015}, a cosmological simulation run with the moving-mesh code \small{AREPO} \citep{Springel2010}.  Illustris was run on a periodic box 106.5 Mpc per side, with a target gas cell mass $m_{\rm{gas}}=1.26 \times 10^6\, {\mathrm M_\odot}$ and dark matter particle mass $m_{\rm{DM}}=6.26 \times 10^6\, {\mathrm M_\odot}$, using a standard $\Lambda$CDM cosmology: $\Omega_{m,0}=0.2726$, $\Omega_{\Lambda,0}=0.7274$, $\Omega_{b,0}=0.0456$, $\sigma_8=0.809$, $n_s=0.963$, and $H_0=70.4 \, \rm{km}\, \rm{s}^{-1} \, \rm{Mpc}^{-1}$ \citep[consistent with][]{Hinshaw2013}.

The simulation includes detailed physics models for galaxy formation and evolution, including primordial and metal-line cooling using a time-dependent UV background \citep{FaucherGiguere2009} and self-shielding \citep{Rahmati2013}; star formation, including associated supernova feedback \citep{SpringelHernquist2003, Springel2005}; and stellar evolution, gas recycling and metal enrichment \citep[see][]{Wiersma2009} incorporating mass and metal loaded outflows \cite[see][]{OppenheimerDave2008, Okamoto2010, PuchweinSpringel2013}.  For a more complete description of the physics models incorporated into these simulations, see \citet{Vogelsberger2014a, Genel2014, Sijacki2015}.

Illustris also directly includes black holes, which are of particular import for this project.  We provide a brief summary of the black hole models here, but see \citet{Sijacki2015} for more complete details.  Black holes are modelled as collisionless sink particles, seeded into any halo with mass $M_{\rm{halo}} > 5 \times 10^{10}\, h^{-1} {\mathrm M_\odot}$ (unless already containing a black hole), with an initial black hole seed mass of $M_{\rm{seed}}=10^5 h^{-1} {\mathrm M_\odot}$.  This model is loosely modelled after a direct collapse formation model \citep[see e.g.][]{HaehneltRees1993, LoebRasio1994, BrommLoeb2003, Begelman2006, ReganHaehnelt2009}, while also intended to remain consistent with a lighter seed model which is followed by efficient mass growth.  Black holes within the simulation grow at a Bondi-Hoyle-like accretion rate set by 
\begin{equation}                                                                                         
\dot{M}_{\rm{BH}} = \frac{4 \pi \alpha G^2 \rho M_{\rm{BH}}^2}{c_s^3}                                    
\label{eq:accretion}
\end{equation}
\citep[see][]{HoyleLyttleton1939, BondiHoyle1944, Bondi1952}, with an imposed upper limit of the Eddington rate 

\begin{equation}                                                                                         
\dot{M}_{\rm{Edd}}=\frac{(4 \pi G M_{\rm{BH}} m_p)}{(\epsilon_r \sigma_T c)}.
\label{eq:eddington}
\end{equation}
Here $M_{\rm{BH}}$ is the black hole mass, $\rho$ is the local gas density around the black hole, $c_s$ is the sound speed of the gas around the black hole, $m_p$ is the proton mass, $\sigma_T$ is the Thompson cross-section, $c$ is the speed of light, and $\epsilon_r$ is the radiative efficiency of the black hole.  During accretion, three modes of black hole feedback are included: ``quasar" and ``radio" modes, which deposit energy to the surrounding gas based on accretion efficiency, and ``radiative" mode, which modifies the nearby gas cooling rate.  

In addition to gas accretion, black holes can grow by merging with other black holes.  Black holes are instantaneously merged when one black hole comes within the smoothing length of another black hole, at which point the two black holes are assumed to merge.  For each merger, the precise time and the masses of the two black holes involved are saved, providing complete merger trees with precise information for the entire population of black hole mergers.  However, we note two significant approximations regarding the timing of these mergers.  First, the simulation incorporates a repositioning scheme to bring black holes to the gravitational potential minimum of the galaxy. This is intended to avoid numerical `wandering' of black holes due to two-body interactions with dark matter or stellar particles, but it also means that each black hole follows a numerical (rather than physical) model to reach the galaxy center (or potential minimum). Secondly, a black hole pair will merge immediately upon coming within one another's smoothing length, without allowing for the possibility of a high-velocity flyby (as the recentering scheme means black hole velocities are not well defined), or incorporating a binary coalescence time.  
As such, black hole mergers within the Illustris simulation are likely to occur more quickly than in the real universe, as the infall to the galactic center is not physically modeled, and the possibility of an initial flyby and the final binary dynamics the are both neglected (and unresolved).  In this project, we consider the impact that a delay in the merger time has on the BH population, and in particular the potential implications for the GW signals of supermassive black hole merger events for LISA.

\subsection{Post Processing}
\label{sec:postprocessing}

\subsubsection{Merger delay}

To investigate how the merger timescale impacts the merging population and the expected detection rates, we calculate a delay time based on the expected dynamical friction in the host galaxies of the massive black holes.  For the sake of this analysis, we focus on how incorporating a more realistic time delay can affect the massive mergers rates from a cosmological simulation \citep[similar to recent work by][]{Katz2020, Volonteri2020}.  As such, rather than computing dynamical friction on the fly during the simulation \citep[as in][for example]{Chen2021}, we take a similar approach to that used by \citet{Volonteri2020} when post-processing the HorizonAGN and NewHorizon simulations, and estimate the dynamical friction timescale for a point mass travelling through an isothermal sphere \citep{Krolik2019}
\begin{equation}
t_{df} = 0.67 \text{ Gyr} \left(\frac{a}{4 \text{ kpc}}\right)^2 \left(\frac{\sigma}{100 \text{ km s}^{-1}}\right) \left(\frac{M_{\text{BH}}}{10^8 M_{\odot}}\right)^{-1} \left(\frac{1}{\Lambda}\right)
\label{eq:delay}
\end{equation}
where $M_\text{BH}$ is the mass of the
infalling black hole, $a$ is the distance to the galaxy center where the central black hole is located, $\Lambda = ln(1+M_{gal}/M_{\text{BH}})$ and $\sigma$ is the central stellar velocity dispersion (approximated as $\sigma = (0.25GM_{gal}/R_{eff})^{1/2}$, where $M_{gal}$ is the stellar mass of the galaxy and $R_{eff}$ is the effective radius of the galaxy).

We use Equation \ref{eq:delay} to calculate the time delay due to dynamical friction, thereby estimating the time it takes for the smaller black hole to reach the larger black hole as it falls into the center of its galaxy.  All properties of the host galaxy and black holes are extracted from the last snapshot before the pair of black holes merge in the original simulation.  We estimate the infall distance ($a$) in two difference ways: one where we assume the infall is the full radius of the galaxy, and one where we set $a$ to the separation distance between the two black holes during that snapshot, with both methods providing qualitatively similar results.  This is a very rough approximation, as it assumes the BH is travelling through a region that could be approximated as an isothermal sphere.  Additionally, for each merger we calculate the dynamical friction timescale at a single instant in time (the snapshot immediately prior to when the merger occurs in the original simulation), without trying to fully model the infall (e.g. by incorporating dynamical friction directly into the simulation).  Nonetheless, it gives us a physically motivated estimate for the additional merging time which should be incorporated, and although the dynamical friction time taken from this instantaneous approximation may be an overestimate (since Equation \ref{eq:delay} is based on spherical orbits, rather than an initial velocity vector for the black hole which may provide an initial infall velocity), we recognize that the true time delay should also include additional binary dynamics, like e.g., loss cone scattering at even smaller scales before the BHs reach coalescence, which would add additional delay times.
\begin{figure}
    \centering
    \includegraphics[width=8.5cm]{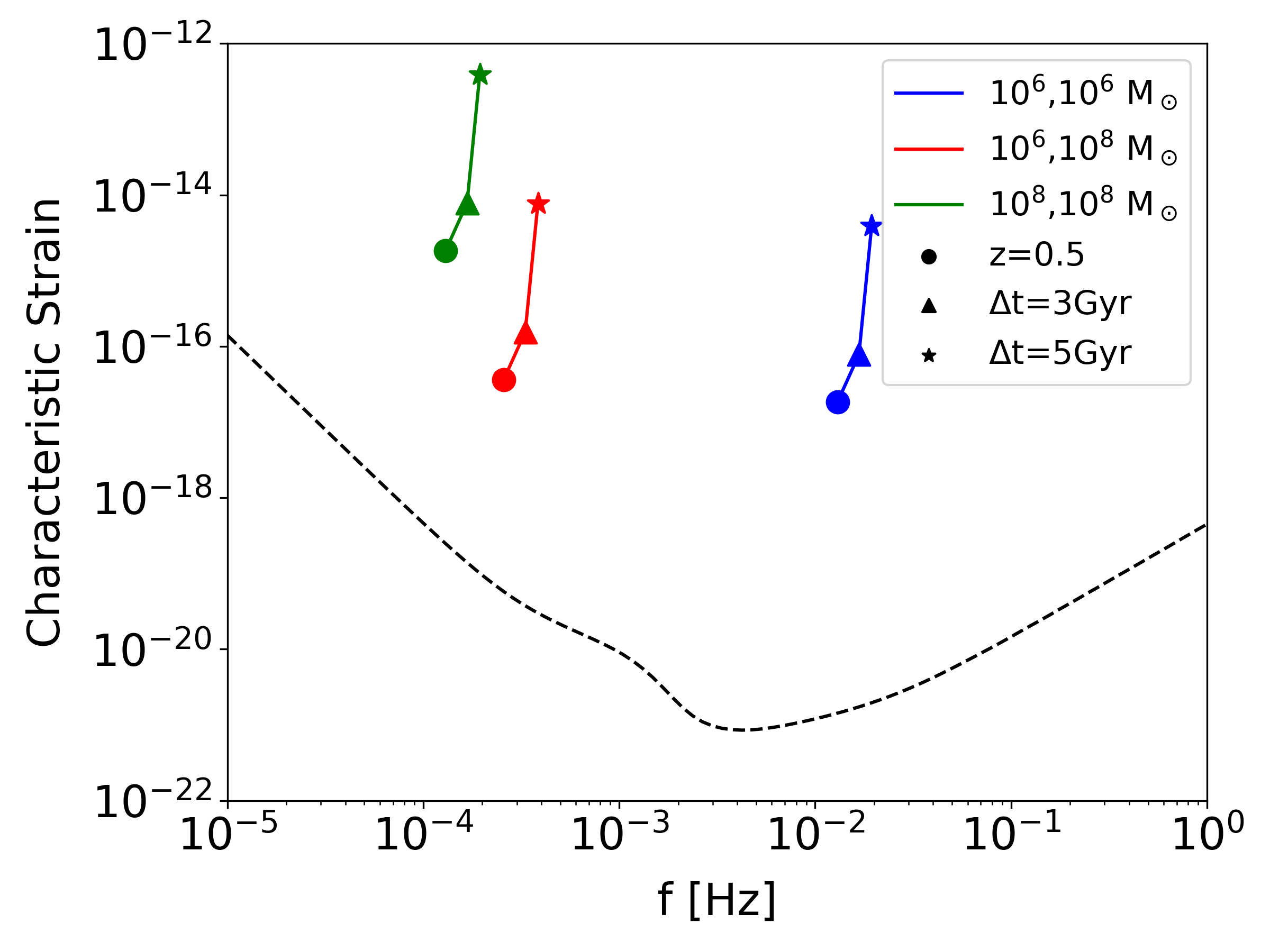}
    \caption{The frequency and strain of the gravitational wave signals emitted by characteristic black hole mergers with $M_1=10^6 M_\odot$, $M_2=10^6 M_\odot$ (blue); $M_1=10^8 M_\odot$, $M_2=10^6 M_\odot$ (red), $M_1=10^8 M_\odot$, $M_2=10^8 M_\odot$ (green), for mergers at redshift 0.5 (circle), and if the merger is delayed by 3Gyr (triangle) or 5Gyr (asterisk), compared to the LISA sensitivity curve (dashed line). This illustrates how  the frequency is primarily determined by $M_1$ (the primary black hole mass), while the strain is more strongly influenced by the secondary mass ($M_2$) and the redshift at which the merger occurs.}
    \label{fig:sample_GWs}
\end{figure}

\begin{figure*}
    \centering
    \includegraphics[width=17cm]{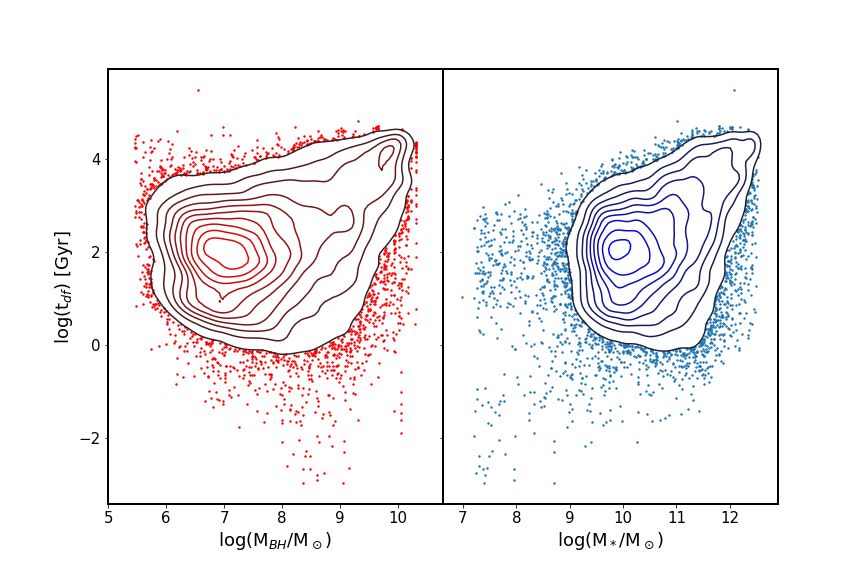}
    \caption{Delay time for Illustris black hole binaries due to the estimated dynamical friction in Eq.(3), shown as a function of black hole mass (left) and host galaxy stellar mass (right). There is substantial scatter among the dynamical friction times at all mass scales, but the highest mass objects tend to have the longest $t_{df}$.}
    \label{fig:delay_vs_mass}
\end{figure*}

The main focus of this work is to study the effects of the delays (induced by e.g. dynamical friction) onto the strain-frequency plane for the GW signal for these massive black hole mergers.
For each merger, we therefore estimate the frequency and strain of the gravitational waves produced using
\begin{equation}
    f_c = 3.9\left(\frac{M}{10^4M_{\odot}}\right)^{-1}(1+z)^{-1}\text{Hz}
    \label{eq:frequency}
\end{equation}
\begin{equation}
    h_{s, circ}(f_c) = \frac{8}{10^{1/2}}\frac{(G\mu)^{5/3}}{c^4d_L}(2\pi f_c)^{2/3}
    \label{eq:strain}
\end{equation}
\citep{Sesana2008}, where $d_L$ is the luminosity distance to the merger, $z$ is the redshift of the merger, $M = M_1 + M_2$ is the sum of the masses of the two merging black holes, and $\mu = \frac{(M_1M_2)^{3/5}}{M^{1/5}}$ is the chirp mass (where $M_2 < M_1$ are the masses of the merging BHs). 

As seen in Equations \ref{eq:frequency} and \ref{eq:strain}, both frequency and strain signals depend on the redshift at which the merger occurs. Thus we expect that adding a time delay which necessarily
changes the redshift at which the merger occurs will also change the frequency and strain of the emitted gravitational wave event, even if all other factors are held constant.
In addition, we expect the merging BH masses prior to the merger will continue to grow by accretion of gas during their inspiral. Adding a delay will ultimately shift $M_1$ and $M_2$ to larger values, which will further affect the strain and frequency of the GW event.
In Figure \ref{fig:sample_GWs} we show the expected GW signal for three sample mergers (ranging from $10^6 M_\odot$ to $10^8 M_\odot$) at z=0.5, as well as the signal if each merger were delayed by 3 Gyr (triangle symbols) or 5 Gyr (asterisk symbols).  Here we see that incorporating a delay on the order of Gyrs will increase both frequency and strain by a significant amount, and that larger black holes masses will push the signal to lower frequencies.

\subsubsection{Mass growth}
\label{sec:massgrowth}
In addition to delaying the redshift/time at which each merger occurs, the delay will also provide more time for the black holes to accrete and grow prior to the merger. As such, one  expects the BH merger masses (once the merger occurs) to be larger than in the original simulation.  The precise amount of accretion will depend on the evolving properties of the local gas, integrated over the full time window.  A fully self-consistent calculation 
would account for BH accretion as the secondary binary inspirals to the central region of the primary black hole, while both accrete gas from their local surroundings. This
would require re-running the simulation, however, which is beyond the scope of this work.  Rather, for this analysis we test two assumptions regarding the effects of accretion onto the BHs when we apply the dynamical friction delay: a Bondi-scaling (i.e. $\dot{M} \propto M_{\rm{BH}}^2$), and an Eddington-scaling (i.e. $\dot{M} \propto M_{\rm{BH}}$). In particular, in each of these two cases (described in more detail below), we assume that the accretion follows a constant scaling from the time of the merger in the original simulation until $t_{\rm{df}}$ after (i.e. when we predict the merger to take place after the dynamical-friction based time delay) and calculate the time-averaged accretion based on the growth of the post-merged black hole from the original simulation.

For the Bondi-scaling case, we assume that $\dot{M}_{\rm{BH}} = \alpha_B M_{\rm{BH}}^2$ (growth proportional to mass squared, as in Bondi-Hoyle formalism, see Equation \ref{eq:accretion}), while in the Eddington-scaling case we assume $\dot{M_{\rm{BH}}} = \alpha_E M_{\rm{BH}}$ (growth proportional to mass, as in the Eddington limit, see Equation \ref{eq:eddington}).  We then use the post-merged mass immediately following the merger in the original simulation (i.e. $M_1+M_2$), and the mass a time $t_{\rm{df}}$ later to determine $\alpha$. In this way we are calculating an effective $\alpha$, equivalent to some efficiency factor assuming accretion conditions remain constant during the time delay.  In the event that the $t_{df}$ window includes one or more additional mergers in the original simulation, we calculate the $\alpha$ `efficiency factor' separately for each successive time window (until the next merger); in this way, we calculate the typical `efficiency' factors based only on gas accretion, not the mass gained by swallowing other black holes.

We expect the Bondi-scaling and Eddington-scaling calculations to provide reasonable upper and lower limits for the true growth rates.  The Eddington-scaling is likely an over-estimate, since it assumes the accretion is proportional to $M_{\rm{BH}}$; we would only expect this to be the case for extended periods of Eddington-limited accretion, and thus in most cases this represents only an upper limit.  Conversely, the Bondi-scaling case assumes accretion is proportional to $M_{\rm{BH}}^2$, which is likely closer to what the simulation would predict (since BHs do not tend to be in an Eddington-limited state for most of the time).  However, it assumes that the local gas environment (representing the gas reservoir for accretion) is the same as it was in the original simulation, which neglects the effect of accretion itself and associated BH feeback.  Since the pre-merger BHs have lower masses than the post-merged BH in the original simulation, we expect the feedback energy liberated in the Bondi-scaling case to be lower than in the original simulation. The weaker feedback would lead to more efficient growth, hence making the Bondi-scaling case likely a lower estimate of the BH mass growth.  Thus, in this project we calculate both the Bondi- and Eddington-scaling cases, which are used to bound the expected range for each individual merger's newly estimated merger masses.

\section{Merging black hole populations}
\label{sec:populations}
\subsection{Merging delays}

\begin{figure}
    \centering
    \includegraphics[width=8.5cm]{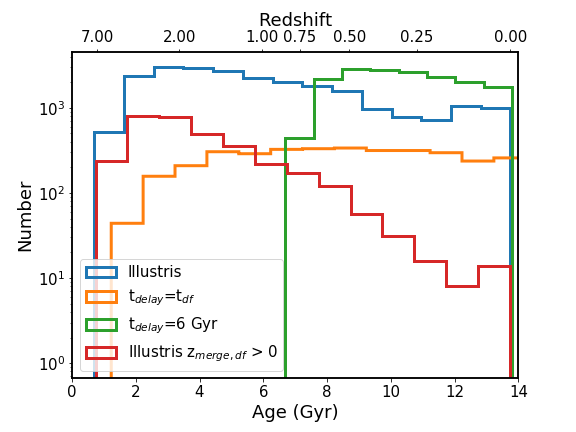}
    \caption{Redshift distribution of mergers.  \textit{Blue:} redshift distribution from the original Illustris simulation.  \textit{Green:} redshift distribution after assuming a fixed 6 Gyr delay.  \textit{ Orange:} redshift distribution after imposing a time delay based on the dynamical friction timescale.  \textit{Red:} original Illustris redshifts, limited to subsample of mergers which are expected to merge before $z=0$, after imposing a dynamical friction time delay. }
    \label{fig:z_distribution}
\end{figure}

In Figure \ref{fig:delay_vs_mass} we plot the expected time delay from dynamical friction (Equation \ref{eq:delay}, as discussed in Section \ref{sec:postprocessing}) as a function of the mass of the black hole (left) and the stellar mass of the host galaxy (right). At higher masses ($M_* \sim 10^{11} M_\odot$) both the average and intrinsic scatter for the expected infall time driven by dynamical friction increases, and the estimated dynamical-friction infall times are long relative to the Hubble time, similar to the findings of \citet{Volonteri2020}.  Delays comparable to and longer than the Hubble time can be expected to have a significant impact on the merger distribution, which we investigate in detail throughout the remainder of this paper.

\subsection{Merger redshifts}
\label{sec:merger_redshifts}

In Figure \ref{fig:z_distribution} we plot the distribution of mergers as a function of cosmic age (lower x-axis) / redshift (upper x-axis) from the original simulation (blue), after applying a fixed 6 Gyr time delay (green), and after applying the dynamical friction estimates for time delay (orange).  We clearly see that the number of black hole mergers in the original simulation peaks at $z \sim 2$.  However, the dynamical friction delay results in far fewer mergers, due to many mergers being delayed past $z=0$ and thus not merging within the current age of the universe.  In fact, we find only $\sim 14\%$ of mergers from the original simulation would be expected to merge after the imposed dynamical friction time delay.  In addition, we see that the distribution of mergers with a dynamical-friction delay (orange) peaks at $z \sim 0.5$, whereas that same sample of mergers in the original simulation (red) peaks much earlier ($z \sim 3$), as the time delay shifts the majority of mergers to much later times.  Hence we expect that incorporating additional physics in upcoming simulations should not only delay mergers to slightly lower redshift, but will also qualitatively change the merger population to peak at significantly lower redshift.

\subsection{Merging masses}

\begin{figure*}
    \centering
    \includegraphics[width=17cm]{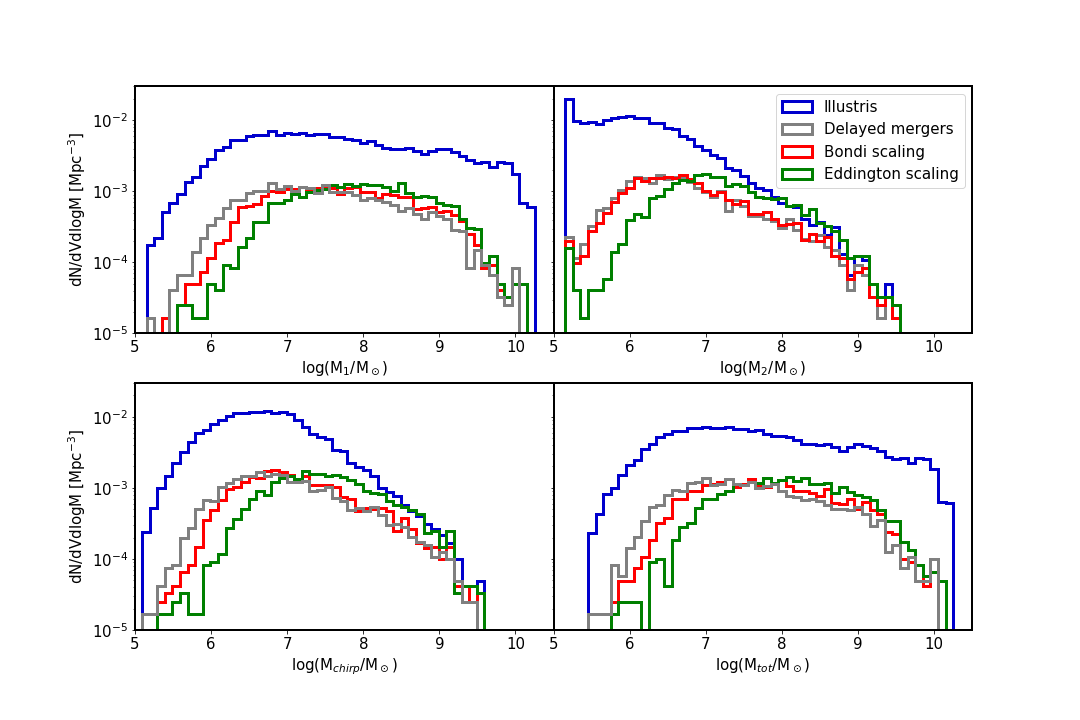}
    \caption{Mass distribution of mergers, for $M_1$, $M_2$, $M_{\rm{chirp}}$, and $M_{\rm{tot}}$ (top left, top right, bottom left, bottom right, respectively.  \textit{Blue:} Original Illustris simulation.  \textit{Grey:} Subsample of Illustris mergers which are expected to merge prior to $z=0$, after imposing a dynamical friction time delay.  \textit{Red:} Distribution of masses after imposing dynamical friction time delay, assuming growth follows a Bondi scaling.  \textit{Green:} Distribution of masses after imposing dynamical friction time delay, assuming growth follows an Eddington scaling.}
    \label{fig:mass_histograms}
\end{figure*}

Because the dynamical friction estimates suggest a significant delay to any given BH merger, we also expect the infalling black holes to continue to grow (as discussed in Section \ref{sec:massgrowth}), leading to higher merging masses than was predicted in the original simulation.  We investigate this by plotting the distribution of merger BHs in Figure \ref{fig:mass_histograms}.  In the top left panel, we plot the distribution of $M_1$ (defined as the more massive black hole involved in a given merger). In blue we see the distribution from the original simulation, while in grey we show the mass distribution of mergers from the original simulation which we still expect to merge before redshift zero after incorporating a time delay (but keeping the original merger masses).  This population of mergers represents $\sim 14\%$ of the total Illustris mergers, and shows a broadly similar distribution as the original simulation, except with fewer high-$M_1$ mergers (note this is before incorporating additional mass growth).  The decrease in high-$M_1$ mergers is expected, since high-masses tend to have long delay times (see Figure \ref{fig:delay_vs_mass}), and also tend to occur at relatively low-z, making them more likely to have their dynamical friction time delay the merger past the current age of the universe.

We compare these merging masses from the original simulation to those after estimating the post-delay merging masses, assuming the infalling BHs grow at the same Bondi scaling (red) or Eddington scaling (green) as the post-merged BH from the original simulation (see Section \ref{sec:postprocessing} for details).  By necessity we see that the added pre-merger growth shifts the typical $M_1$ to higher masses, but we note that this primarily occurs among low-mass mergers, assuming Eddington-scaled growth.  In particular, at $M_1 \sim 10^7 M_\odot$ the typical infalling BH will grow by a factor of $\sim 2$ under assumed Bondi scaling, and a factor of $\sim 7$ under assumed Eddington scaling (see Table \ref{tab:typicalmasses} for mean and scatter of mass distributions).  In contrast, for $M_1 > 10^9 M_\odot$ the typical growth is only about 15\% (Bondi) or 23\% (Eddington), since most high-mass black holes tend to have more suppressed growth. Additionally, we note that the $M_1$ growth is relatively insensitive to the growth model (Bondi vs. Eddington). This is due to the high $M_1$ mergers tending to have a very high mass ratio: a high mass ratio merger means $M_1$ is impacted relatively trivially when swallowing $M_2$, and thus there is only a small impact on the total growth.  In contrast, low-$M_1$ mergers generally have high mass ratios, hence delaying the merger more significantly impacts the black hole mass, and the resulting growth is more sensitive to an accretion efficiency proportional to $M_{\rm{BH}}$ (Eddington) or $M_{\rm{BH}}^2$ (Bondi).  

We also plot the distribution of $M_2$ (the smaller BH mass in a given merger), $M_{\rm{chirp}}$ (the chirp mass for the merger), and $M_{\rm{tot}} \equiv M_1 + M_2$ (the total merger mass) in the top right, bottom left, and bottom right panels, respectively. As in the $M_1$ panel, we see that the merging masses are affected most strongly at low-masses, with more effective Eddington-scaled growth having the strongest effect.  We also note that the  Bondi-scaled growth has minimal impact on $M_2$, which we would expect given that $M_2$ is generally much smaller than $M_1$: if we assume both $M_1$ and $M_2$ grow following the same Bondi scaling as the post-merged BH in the original simulation, then $M_2$ will grow very little since $M_2 << M_1 + M_2$ and Bondi accretion scales like $M_{\rm{BH}}^2$.

\begin{table*}
    \centering
    \begin{tabular}{c|c|c|c|c|c|c|c|c|c|c|c|c|c|c|c}
        \hline
    
        BH Sample & \multicolumn{4}{c}{$<\rm{log}(M/M_\odot)>$} & & \multicolumn{4}{c}{$\sigma (\rm{log}(M/M_\odot))$}  \\
        \cline{2-5} \cline{7-10}
         & $M_1$ & $M_2$ & $M_{\rm{chirp}}$ & $M_{\rm{binary}}$ & & $M_1$ & $M_2$ & $M_{\rm{chirp}}$ & $M_{\rm{binary}}$\\
        \hline
        Base Population             & 7.7 & 6.1 & 6.7 & 7.7 & & 1.1 & 0.7 & 0.7 & 1.1  \\
        Delayed mergers             & 7.5 & 6.7 & 7.0 & 7.6 & & 0.9 & 0.8 & 0.8 & 0.8 \\
        Bondi-scaled growth     & 7.7 & 6.8 & 7.1 & 7.8 & & 0.9 & 0.8 & 0.7 & 0.8 \\
        Eddington-scaled growth & 8.0 & 7.2 & 7.5 & 8.1 & & 0.8 & 0.7 & 0.7 & 0.8 \\

         \hline
    \end{tabular}
    \caption{Mean and standard deviation of each population}
    \label{tab:typicalmasses}
\end{table*}

\begin{figure}
    \centering
    \includegraphics[width=8.5cm]{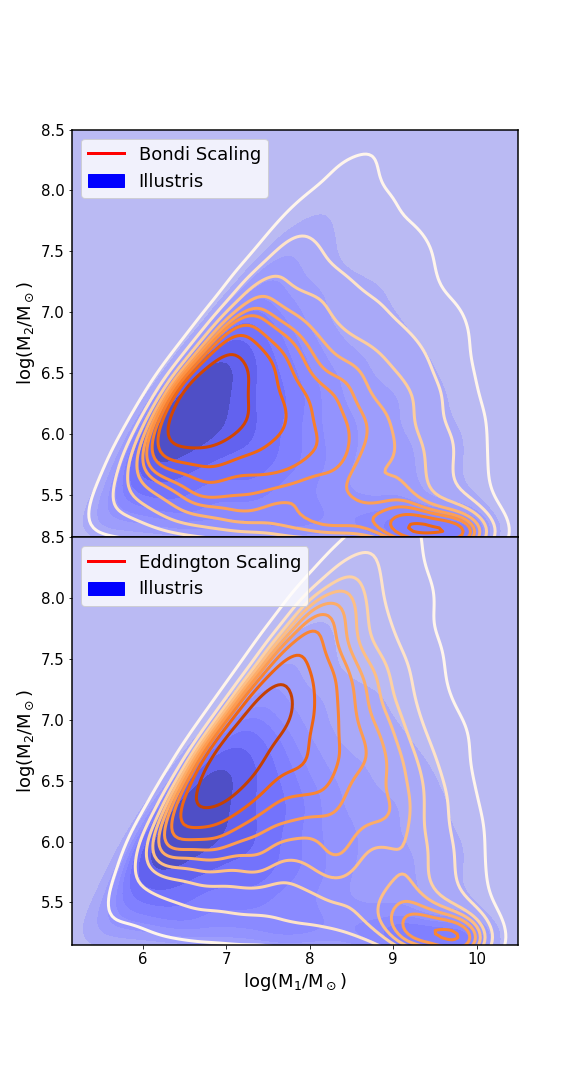}
    \caption{\textit{Top:} Distribution of merger masses ($M_1$ and $M_2$), from the original Illustris simulation (blue) and after mass growth during the merger delay time, assuming growth follows a Bondi scaling (orange contours). \textit{Bottom:} Same as top, but assuming growth follows an Eddington scaling.}
    \label{fig:M1M2}
\end{figure}

To more clearly visualize how the delay-time growth impacts merging masses, in Figure \ref{fig:M1M2} we plot the distribution of masses in the $M_1 - M_2$ plane, for the original Illustris data (blue shaded contours) and the expected masses after the DF delay time growth for both Bondi scaling (top) and Eddington scaling (bottom).  In all cases we see a qualitatively similar distribution: a peak at moderate $M_1$ and $M_2$ a factor of a few smaller, and a secondary peak for high $M_1$ and very low $M_2$.  As in Figure \ref{fig:mass_histograms} we see that assuming growth driven by Bondi scaling (top panel) tends to result in relatively minor BH growth (i.e. the merger mass is not substantially changed as a result of the delay), with the largest impact among $M_1 \sim M_2$ mergers.  In contrast, assuming Eddington-scaled growth means that the pre-merged BHs can grow more efficiently, resulting in a more significant increase in merging masses, including among the highest mass ratio mergers (bottom panel).

\begin{figure}
    \centering
    \includegraphics[width=8.5cm]{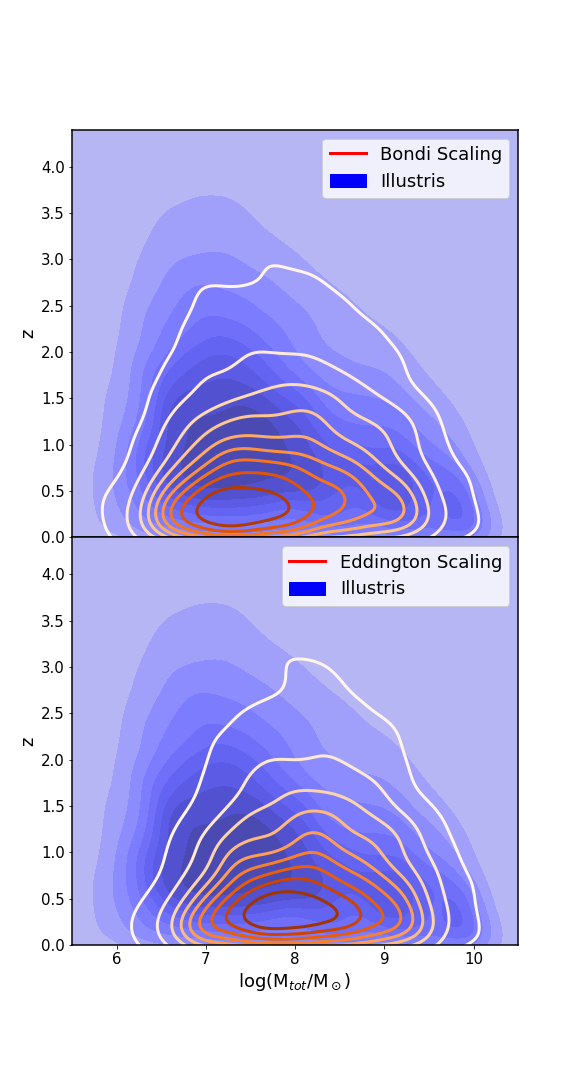}
    \caption{\textit{Top:} Distribution of mergers across redshift and total merger mass ($M_{\rm{tot}} = M_1+M_2$) from the original Illustris simulation (blue), and after imposing a dynamical friction time delay, assuming growth follows a Bondi scaling.  \textit{Bottom:} Same as top, but assuming black hole growth follows an Eddington scaling.}
    \label{fig:zMtot}
\end{figure}

Similarly, in Figure \ref{fig:zMtot} we show the distribution of mergers in mass-redshift space.  Here we again see that the delay significantly shifts the peak of the distribution to lower redshift (recall Figure \ref{fig:z_distribution}), while only Eddington scaled growth tends to cause a significant shift in merger mass.  Additionally, we see that the increase in merging masses is larger at high-redshift.  This is indicative of black holes tending to grow more efficiently at early times, and also suggests that properly modeling the expected merger timescale will be crucial to interpreting high-redshift measurements (e.g. using LISA to constrain black hole seed formation models), since black holes tend to grow more efficiently at early times \citep{DeGrafBHGrowth2012}.  

\begin{figure*}
    \centering
    \includegraphics[width=17 cm]{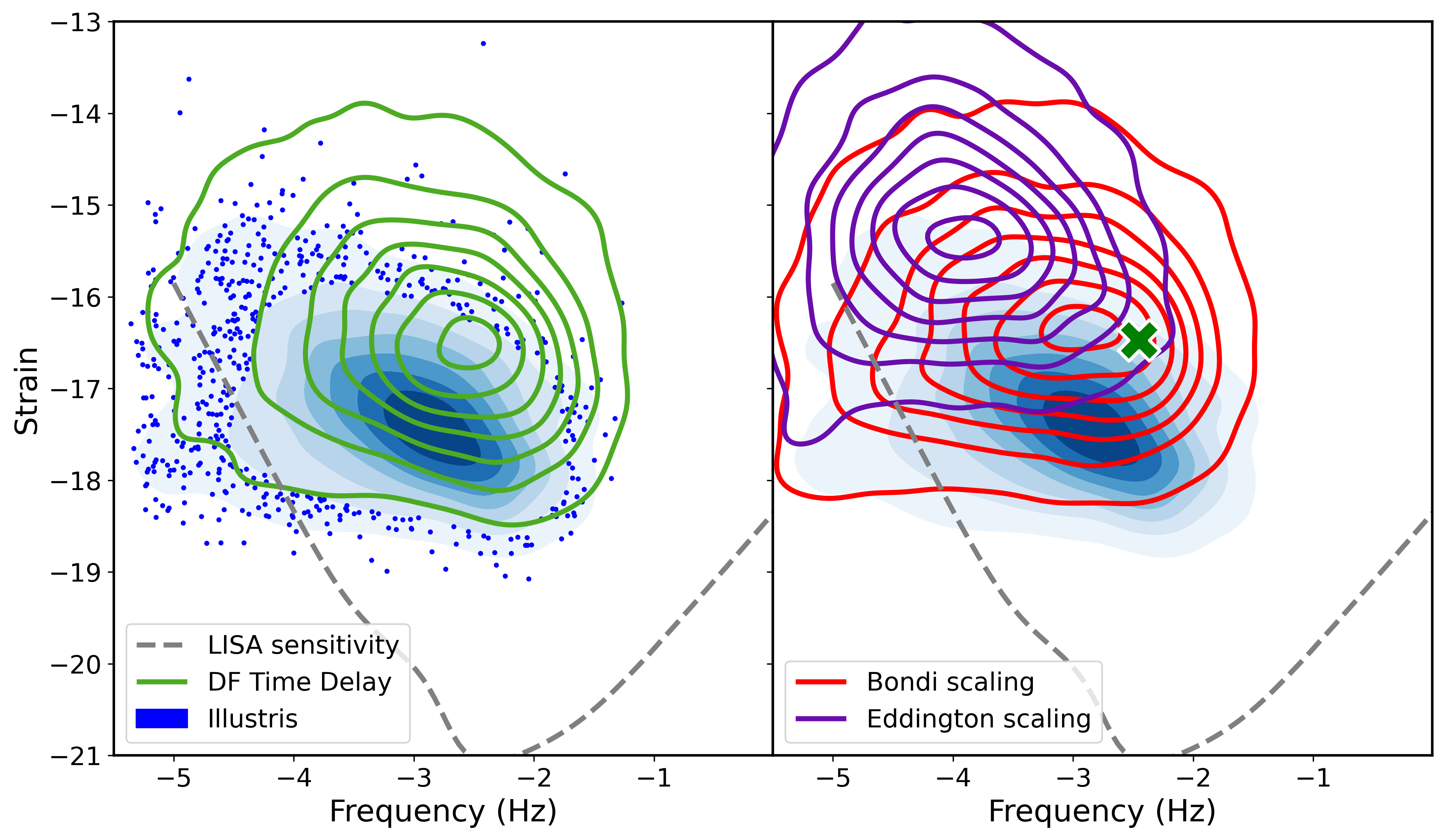}
    \caption{\textit{Left panel:} Distribution of GW signals in frequency-strain space for the original Illustris simulation (blue), and after incorporating a DF time delay (green).  \textit{Right panel:} Distribution of GW signals after incorporating black hole mass growth according to Bondi scaling (red) or Eddington scaling (purple).  The green cross shows the peak of the distribution without any mass growth (i.e. peak of the green contours from left panel).  Incorporating a time delay primarily increases the strain, with a minor increase in frequency.  Incorporating mass growth decreases the frequency and further increases the strain, especially for more efficient growth (i.e. assuming Eddington scaling).}  
    \label{fig:freq-strain}
\end{figure*}

\section{GW detectability}
\label{sec:GWdetect}

Having shown that incorporating a time delay based on dynamical friction can significantly alter the redshift and masses of the full merger population, we now consider the impact these delays can have on the detectability of the gravitational wave strain for these events.  Similar to Figure \ref{fig:sample_GWs}, we calculate the expected frequency and strain of the emitted gravitational waves using Equations \ref{eq:frequency} \& \ref{eq:strain}, and plot the results in the left hand panel of Figure \ref{fig:freq-strain}, together with the LISA sensitivity curve \citep[dashed grey line, see][]{Robson2019}. In the blue shaded contours we show the frequency-strain distribution for the original Illustris simulation, and in the green contours we compare to the distribution after incorporating the dynamical friction time delay, while assuming the merging masses remain unchanged.  
There is a significant impact on the strain and frequency of the events when incorporating this realistic time delay for a black hole merger. If we look at where majority of the events take place (the innermost contours) the peak of the signal is increased by approximately 0.4 dex in frequency, and $\sim$1 dex in strain. A time delay in the BH merging population implies an overall increase in the fraction of mergers that will be within the LISA sensitivity, and with a stronger signal given. However, due to the added time delays, most mergers from the original simulation will actually not occur: the addition of the dynamical friction infall time delays the majority of mergers until they no longer take place within the current age of the universe (as discussed in Section \ref{sec:merger_redshifts}).  We also note that incorporating a time delay does not significantly affect the scatter in frequency ($\sigma=0.77$ dex in the original simulation, and 0.78 dex after including a time delay), but does increase the scatter in the predicted strain amplitude (from $\sigma=$ 0.73 to 0.94 dex).

In the right hand panel of Figure \ref{fig:freq-strain}, we show the impact of 
the merger timescale after including the associated mass growth (as discussed in Section \ref{sec:postprocessing}) in addition to the delayed merger redshift. We show the results of Bondi-scaled mass growth with red contours, and the Eddington-scaled mass growth with purple contours.  As in the left hand panel, we show the distribution from the original simulation (blue shaded contours), and for comparison we show the peak of the distribution  a dynamical friction time delay but no mass growth (i.e. the peak of the green contours in the left panel) as a green X.
Again we see a significant effect when including both time delay and mass growth, with a comparable shift in the peak of the distributions.  Although the Bondi-scaled mass growth has a relatively minor impact on the location of the peak of the frequency-strain distribution, it does increase the scatter in frequency (from 0.78 dex to 0.85 dex), largely dominated by the low-frequency tail. 

\begin{table*}
    \centering
    \begin{tabular}{c|c|c|c|c}
        \hline

        BH Sample & $<\rm{log (f/Hz)}>$ & $\sigma (\rm{log (f/Hz)})$ & $<\rm{log (h)}>$ & $\sigma (\rm{log (h)})$ \\
        \hline
        Base Population             & -3.1 & 0.77 & -17.2 & 0.73 \\
        Delayed mergers             & -2.9 & 0.78 & -16.4 & 0.94 \\
        Bondi-scaled growth     & -3.4 & 0.85 & -16.3 & 0.95 \\
        Eddington-scaled growth & -4.0 & 0.69 & -15.3 & 0.94 \\
         \hline
    \end{tabular}
    \caption{Mean and standard deviation of frequency and strain for GW signals, for the original Illustris simulation, after incorporating a time delay, and after incorporating additional Bondi-scaled and Eddington-scaled black hole mass growth.}
    \label{tab:freq_strain}
\end{table*}

In contrast, the Eddington-scaled growth case substantially decreases the peak frequency and increases the peak strain (see also Table \ref{tab:freq_strain}).  Furthermore, the 1-$\sigma$ scatter in frequency decreases (from 0.78 dex to 0.69 dex), while the scatter in strain remains unchanged (at 0.94 dex).  From Equation \ref{eq:frequency} we see that $f_c \propto 1/(M_1 + M_2)$, hence we find that allowing the black holes to grow during the merging time delay leads to decreased GW frequency, and the range of black hole growth rates during that time results in increased scatter in $f_c$ signal.  On the other hand, we see in Equation \ref{eq:strain} that $h \propto (M_1 M_2)^{3/5}/(M_1+M_2)^{13/15}$, hence there is very little impact on the strain's scatter.  

\begin{figure}
    \centering
    \includegraphics[width=8.5cm]{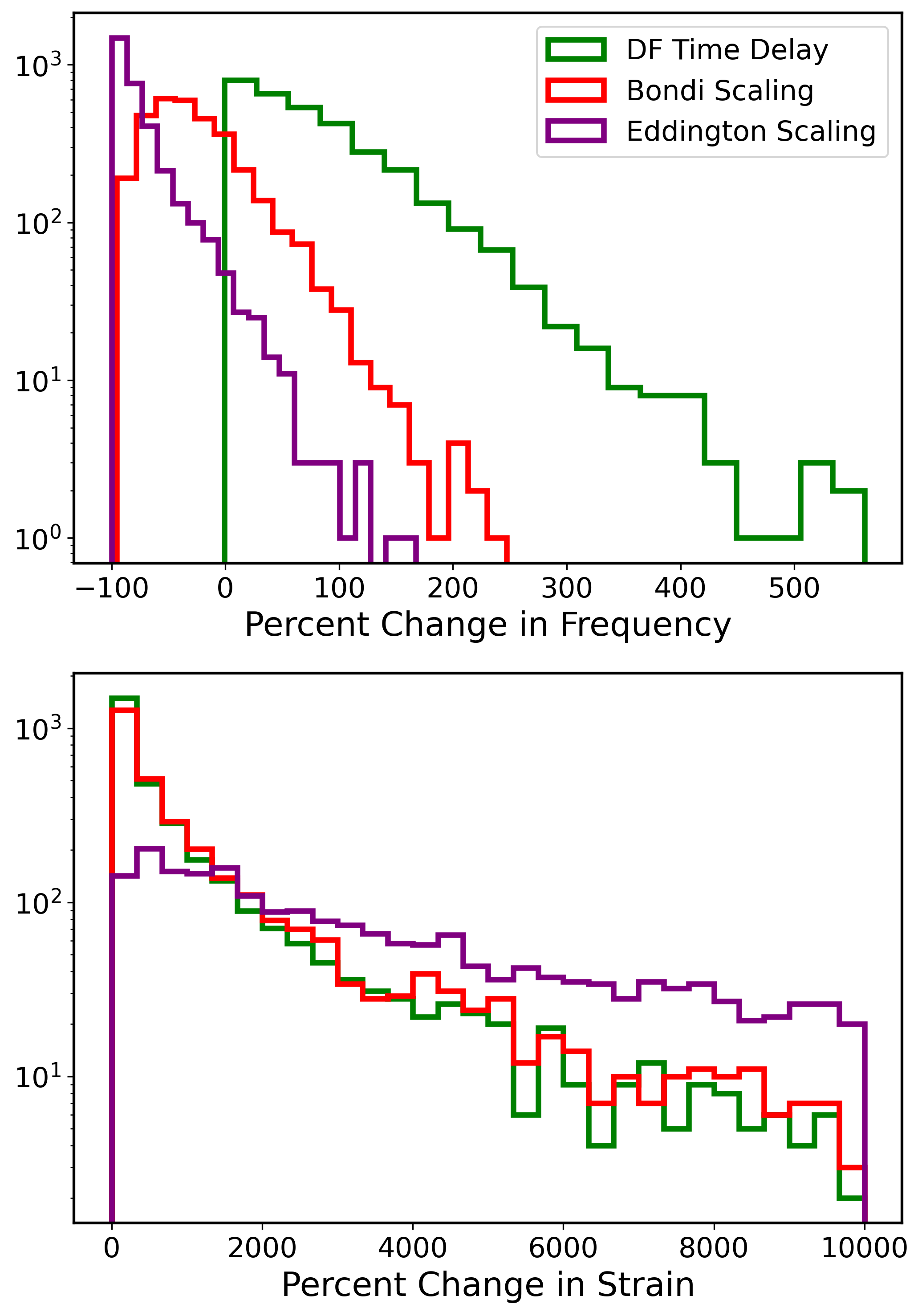}
    \caption{Distribution in the fractional change in frequency (top) and strain (bottom) of the gravitational wave signal.  As in Figure \ref{fig:freq-strain}, we see that a time delay increases both frequency and strain, while the mass growth further increases the strain but decreases the frequency, generally resulting in a smaller frequency than that from the original simulation. } 
    \label{fig:GWsignal_change}
\end{figure}

In Figure \ref{fig:GWsignal_change} we plot the fractional change in frequency (top) and strain (bottom). In green we show the fractional change in frequency and strain caused by the the dynamical friction time delay without incorporating mass growth, while in red (purple) we show the fraction change caused by the use of both the time delay and the Bondi- (Eddington-) scaled mass growth.  As in Figure \ref{fig:freq-strain}, we see that the time delay (green) increases both frequency and strain. When the appropriate mass growth is also included (red and purple), the majority of GW signals decrease in frequency, but some mergers nonetheless have a net increase in frequency ($\sim 20\%$ of mergers assuming Bondi-scaled growth, and $\sim 5\%$ for Eddington-scaled growth).  We also see that incorporating a dynamical friction time delay results in a substantial increase in GW strain amplitude, which is further increased by incorporating Eddington-scaled growth (and as in Figure \ref{fig:freq-strain} we see that Bondi-scaled growth has relatively minimal impact on the strain of the GW signals).

\break
\subsection{Detectability rate}
\begin{figure}
    \centering
    \includegraphics[width=8.5cm]{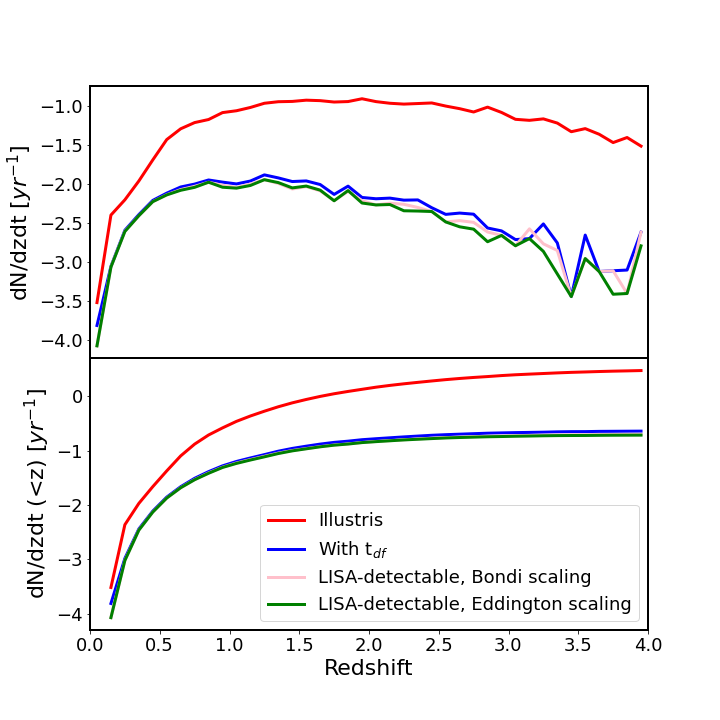}
    \caption{Expected GW signal rate as a function of redshift (top) and cumulative with redshift (bottom).  Red - Signal rate based on original Illustris simulation; Blue - Signal rate after incorporating time delay; Green/Pink - Signal rate of LISA-detectable mergers, assuming Eddington/Bondi scaled growth. Incorporating a time delay has a strong impact on the expected GW rate, with the majority of such mergers being detectable by LISA (though note the rate is only for the high-mass end of the LISA sensitivity range which is probed by Illustris, see text for details).}
    \label{fig:avgz_detect}
\end{figure}

We also use the full population of BH mergers to predict the global rate of GW events reaching Earth, as a function of merger redshift, i.e. integrating the merger rate in the simulation over the cosmic volume at each redshift: 
\begin{equation}
    \frac{{\rm d}N}{{\rm d}z\,{\rm d}t} = \frac{1}{z_2-z_1} \int_{z_1}^{z_2} \frac{{\rm d}^2 n(z)}{{\rm d}z\,{\rm d}V_c} \frac{{\rm d}z}{{\rm d}t} \frac{{\rm d}V_c}{{\rm d}z} \frac{{\rm d}z}{1+z}\,
\end{equation}
We plot the expected event rate from the original Illustris simulation (red) and after including our dynamical friction time delay (blue) in Figure \ref{fig:avgz_detect}, both as a function of redshift (top) and as a cumulative distribution (bottom).
Although the original simulation suggests that $\sim 3$ GW signals should reach the Earth each year (for merging black holes with masses probed / simulated in Illustris, see caveat below), with the effect of the delay for the inspiral time this estimate drops by roughly an order of magnitude.  This decreased rate is due to a combination of high-redshift mergers being delayed to later times (with correspondingly smaller cosmological volumes and thus fewer emitted GW events), and mergers, especially at low redshift, whose infall time delays them beyond the current age of the universe and therefore will not have merged or emitted GWs yet (as discussed in Section \ref{sec:merger_redshifts}).  In addition to the overall rate of GW events, we also see that incorporating a time delay shifts the peak of the GW signal to later times, such that after incorporating a time delay we would expect the peak GW event rate to occur at z $\sim 1.25$, rather than the peak at z $\sim 2$ as found in the original simulation \citep[also comparable to predictions by][]{Volonteri2020, Katz2020}.

We also consider the event rate of GWs
which we expect to be detectable by LISA (as discussed in Section \ref{sec:GWdetect}). The merging masses are important when imposing a sensitivity cut (see Figure \ref{fig:freq-strain}), so we show the results of a sensitivity cut assuming both Bondi- (pink) and Eddington- (green) scaled growth. As we saw in Figure \ref{fig:freq-strain}, the majority of mergers produce GW signals within the detectability range for LISA regardless of the mass growth model, hence in Figure \ref{fig:avgz_detect} we see that the rate of LISA-detectable GWs (pink and green) is comparable to the overall event rate after having imposed a dynamical friction based time delay.
We see the rate of LISA-detectable GW events is nearly identical to the total number of GWs, with only minor changes at high redshifts.  

We note a strong caveat to the rates plotted in Figure \ref{fig:avgz_detect}, which are limited to the mass scales resolved in Illustris.  The Illustris black holes are limited to the high-end of LISA's peak sensitivity, and the majority of LISA detections would be expected to come from mergers below Illustris' black hole seed mass \citep[see, e.g.][]{Ricarte2018, Dayal2019, DeGrafSijacki2020, Volonteri2020}.  As such, the rates plotted should not be interpreted as predictions for all LISA detections, but rather to demonstrate the importance of including a time delay when predicting merger signals, at least for the high-mass end of the LISA-sensitivity range.

\section{Scaling relationship with host galaxy}
\label{sec:scaling}

\begin{figure*}
    \centering
    \includegraphics[width=17cm]{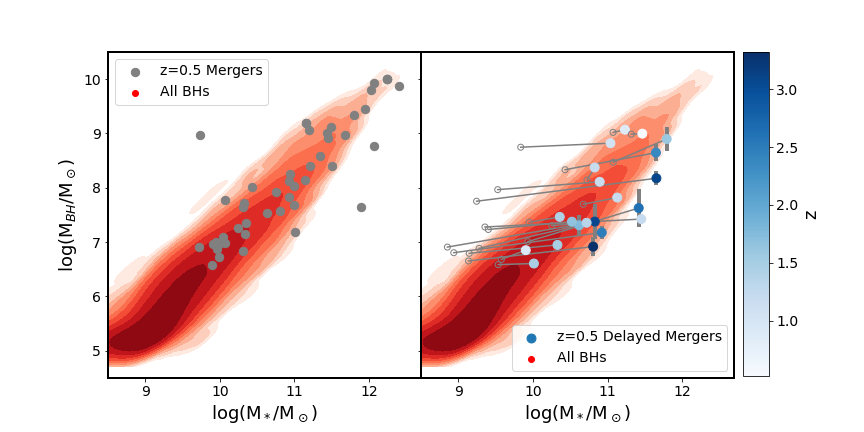}
    \caption{\textit{Left panel:} Contour plot of galaxy stellar mass vs. mass of its central black hole from the original simulation at z=0.5 (red contours), with [$M_{\rm{BH}}-M_*$] of mergers which occur within 25 Myr plotted as grey circles.  \textit{Right panel:} As in the left panel, but the solid circles are based on mergers after incorporating a dynamical-friction based time delay.  The blue points record the average of predicted Eddington and Bondi scaled mass growth, with error bars spanning the range between these two predictions. The circles are colour coded by the redshift of the original mergers (i.e. before the time delay was added). Each blue point is connected to a corresponding grey open circle, which shows the BH and stellar mass of the same merger in the original simulation (i.e. before incorporating the time delay).}
    \label{fig:scaling_both}
\end{figure*}

In addition to considering the impact that merger delays have on the expected merger rates and merging masses, we also note that any delay to black hole mergers will also affect the black hole growth history and thus the masses of the black holes found at the center of galaxies.  To investigate any effect this may have, we consider the black hole - galaxy scaling relation and where the detectable mergers occur on the $M_{\rm{BH}}-M_*$ plane.

Here we focus on $z \sim 0.5$, to highlight a time when the majority of mergers have been significantly shifted to later time (see Figures \ref{fig:z_distribution} \& \ref{fig:zMtot}), and compare the $M_{BH}$-$M_*$ relation of mergers near this time to the underlying scaling relationship.  In Figure \ref{fig:scaling_both} we plot these relations, for the original mergers and after incorporating a time delay.

In red contours we plot the $M_{\rm{BH}}-M_*$ relation, for the full population of galaxies at z=0.5 (where $M_{\rm{BH}}$ is the mass of the most massive black hole in the given galaxy), showing a clearly defined scaling relation.  We compare this to the set of z $\sim 0.5$ mergers, which we define as mergers occurring within 25 Myr of z=0.5.  We then take the total merger mass ($M_1+M_2$) at the time of the merger and the stellar mass of the galaxy hosting the merger (with $M_*$ coming from the z=0.5 snapshot), which we plot in the left hand panel of Figure \ref{fig:scaling_both}.  Here we see that most mergers in Illustris occur in galaxies which lie along the typical scaling relation, and the merger distribution appears roughly uniform.  

In addition to the $z \sim 0.5$ mergers from the original simulation (left hand panel), in the right hand panel we incorporate the dynamical friction time delay.  We again select mergers within 25 Myr of $z \sim 0.5$, but in this case we use the merger time after incorporating the delay (i.e. using $t_{\rm{Illustris,merger}}+t_{df}$, rather than $t_{\rm{Illustris,merger}}$ as in the left hand panel).  For these mergers, we again take the galaxy stellar mass from the z=0.5 snapshot, and we estimate $M_{\rm{BH}}$ under both the Bondi-scaling and Eddington-scaling assumptions (See Section \ref{sec:massgrowth}). The midpoint of between these two bounds are plotted as blue circles, and the $M_{\rm{BH}}$ ranges spanned by these estimates are shown as grey error bars. 

After incorporating the time delay, the merger hosts are still found spanning a wide range of galaxy masses, but we now find that the merging black hole masses (blue circles) tend to be slightly undermassive compared to the overall distribution (red contours).  We perform a linear fit to the merger scaling relation, and confirm that the delayed mergers (solid points in right panel) are indeed less massive relative to their host when compared to the non-delayed mergers (points in left panel): the normalization is $\sim 0.4$ dex lower for the delayed mergers.  However, the slope of the best fitting relation is only $\sim 5\%$ lower for the delayed mergers.  Hence the scaling relation is broadly consistent, except that including delays results in merger masses slightly less massive than a typical non-merging black hole found in an equivalent mass galaxy.

To better understand the cause of that shift, for each merger we also plot (open grey circles) the merger mass and host stellar mass of the merger which was delayed until $z \sim 0.5$, connected to the delayed merger with a grey line.  Furthermore, the blue circles are color-coded by the redshift at which the undelayed merger took place in the original simulation.  Here we see that the un-delayed mergers (which occurred at higher redshift) were in lower mass galaxies (as expected), and were more likely to be overmassive relative to the z=0.5 relation. The delayed mergers which lie below the z=0.5 relation tend to be the mergers which have been delayed the longest (i.e. which merged at the earliest times in the original simulation, represented by darker blue).  This is unsurprising, since a longer delay corresponds to a longer time during which the black holes grow.  Delaying a merger means that the pre-merger accreting black holes have a lower mass than the post-merged BH from the original simulation, leading to a lower accretion rate, and thus the largest difference occurs in those which have the longest time to grow (i.e. the longest delays). We also note that the mergers lying below the typical scaling relation tend to have the largest error bars: this means that those objects have the largest difference between the Eddington- and Bondi-scaled growth, which tends to happen for major mergers (since delaying a 1:1 merger has the biggest impact on the initial mass).  In summary, after incorporating a dynamical-friction estimated merger time the majority of mergers still take place well within the typical scaling relation, but we also find a significant subsample of high-redshift major mergers which end up relatively undermassive relative to their host galaxy which we did not find in the original simulation.

\section{Conclusions}
\label{sec:conclusions}
In this work, we have analyzed the supermassive black hole mergers in the Illustris simulation, and quantified the impact that a realistic merger timescale (estimated based on dynamical friction) will have on the population of black hole mergers, in terms of merger rates, merging masses, gravitational wave signals, and connections to their host galaxies.

\begin{itemize}
\item We estimate the merger delay by a post-processing calculation of the delay due to dynamical friction; this timescale represents a proxy for the infall time for the secondary black hole to reach the galaxy center, where the central BH resides and hence the BH pair can merge. The merger delay timescales estimated in this way tend to be long, often longer than the Hubble time, and hence we expect that merger delays due to dynamical friction in the host to have a significant impact on the overall merger populations. We note this is consistent with recent analysis of the Illustris, HorizonAGN, and NewHorizon simulations \citep[see, e.g.,][]{Katz2020, Volonteri2020}.
\item The longest delays tend to occur for the most massive objects: the most massive black holes tend to be found in the most massive galaxies, which require the longest dynamical friction timescale to fall in to the galaxy center.
\item The long infall times result in many mergers being delayed past the current age of the universe, thereby substantially decreasing the total number of black hole mergers we would expect to occur, and decreasing the rate at which GWs from merging BHs would reach Earth by $\sim$ 1 dex (at least at the mass scales probed by Illustris).
\item Low redshift mergers are most likely to be delayed past z=0 (primarily because they have the least time until z=0, but also because the highest mass galaxies, which tend to have longer dynamical friction times, are found at late times).  As such, black hole merger rates may remain relatively flat from z $\sim 1$ to 0 with a GW signal that peaks at z$\sim 1.25$, in contrast to the original simulation, in which the GW signal is predicted to peak at z$\sim 2$.
\item The delayed mergers provide more time for black holes to grow prior to coalescence, leading to an increase in typical merger masses, ranging from $\sim$ 0.1 dex (if accretion efficiency tends to scales with $M_{\rm{BH}}^2$), to $\sim 0.5$ dex (if accretion efficiency scales with $M_{\rm{BH}}$).  We expect the true values to lie somewhere between these two extremes (i.e. typically between 0.1 - 0.5 dex), emphasizing the importance that future simulations incorporate black hole accretion during the infall/inspiral time.
\item The decrease in z$>$0 black hole mergers leads to a corresponding decrease in the expected GW signal we expect to observe with LISA.  Among the mergers which still occur (i.e. inspiral times short enough to complete before redshift 0), we find the delay tends to increase the GW signals' frequency by $\sim 0.4$ dex and the strain by $\sim$ 1 dex.  Incorporating the expected additional pre-merger growth as well results in a net increase in strain ($\sim 1-2$ dex) and decrease in frequency ($\sim 0-1$ dex), depending on the efficiency of accretion during the time delay.
\item In the original simulation, merging SMBHs tend to lie on the typical $M_{\rm{BH}}-M_*$ relation, implying mergers tend to occur in typical galaxies.  Incorporating the dynamical friction based time delay, however, tends result in merging SMBHs being somewhat undermassive relative to their host galaxy, especially among the mergers with the longest time delays.  Although the time delay affords the black holes more time to grow, they tend to grow less efficiently than if they had merged more quickly (as in the original simulation), while the galaxy in which the infalling BH is found continues to grow efficiently (necessarily un-affected by our post-processing analysis).  The net result is that black holes with comparatively short infall times tend to be found in galaxies on the typical black hole-galaxy scaling relation; on the other hand, those with longer infall timescales tend to end up undermassive relative to the overall (i.e. non-merging) BH population.  This suggests that GW signals detected by LISA and PTAs may include a slight bias toward undermassive black holes when compared to the complete (non-merging) scaling relation.  Our estimates here suggest this bias is relatively weak, but is particularly sensitive to the accretion rates during the infall time.  
\end{itemize}

In summary, we have shown that incorporating a longer merger time (e.g. including a dynamical friction model for black hole infall, and a hardening time for binary coalescence) would be expected to have a significant impact on merging black hole populations: there will be fewer overall mergers, peaking at lower redshift (z $\sim 1.25$ rather than 2) and higher masses (on the order of $\sim 0.1-0.5$ dex), resulting in typical GWs having lower frequency and higher strain than would otherwise be predicted.  The precise magnitude of these shifts are sensitive not only to the calculated delay time, but also the efficiency with which each individual black hole grows during this time (especially the GW signals emitted by these mergers). 

We note that (as discussed above) the results of this paper are sensitive to both the delay timescale for each merger, as well as the mass growth model which takes place during that time.  As such, the precise quantitative effects depend on several models, as has been previously found.  In particular, the delay time is sensitive to the dynamical friction model \citep[see, e.g.][]{Tremmel2015, Tremmel2017, Pfister2019, Sharma2020, Ricarte2021, Chen2021}, as well as the binary coalescence timescale \citep[see, e.g.][]{Kelley2017, Pfister2017, Rantala2017, Volonteri2020}, which can be sensitive to a variety of binary parameters that also impact the GW frequency/strain, including black hole spin and binary eccentricity \citep[e.g., see][]{Kelley2017}.  Furthermore, as shown in Figures \ref{fig:mass_histograms}-\ref{fig:GWsignal_change}, the results are sensitive to the models used for accretion rates during the infall and binary hardening times, for which any post-processing analysis necessarily relies on generalizing assumptions.  Overall, given the importance that merging timescales and associated black hole growth can have on merger populations, including the expected GW signals that will be detected by LISA and PTAs (especially among high-redshift mergers), we emphasize the importance that future simulations incorporate physically motivated models for merger timescales (both infall and binary hardening times) and self-consistently model the black hole growth which occurs during these periods.

\section*{Acknowledgements}
This research has been carried out by a group  of undergraduate students (first six authors of this paper) who started working together as part of their freshman
class "First Years Seminars: Black Hole Mergers and Gravitational Waves" taught in spring 2020 by T.Di Matteo, Physics Department at Carnegie Mellon University. 
As the students expressed interest in continuing their coursework projects,
C. DeGraf met (on zoom) with this group over the last year to supervise the research and bring it to completion with this paper.
TDM acknowledges funding from NSF AST-1616168, NASA ATP  80NSSC20K0519,
 NASA ATP 80NSSC18K101, and NASA ATP NNX17AK56G, and 
This work was also supported by the NSF AI Institute: Physics of the Future, NSF PHY-2020295.

\section*{Data Availability}
The data underlying this article were derived from sources in the public domain: The Illustris Project (https://www.illustris-project.org/).

\bibliographystyle{mn2e}       
\bibliography{astrobibl}       

\end{document}